\documentclass[sigplan, screen, authorversion, nonacm]{acmart}
\AtBeginDocument{%
  }
\usepackage[xcdraw]{xcolor}
\usepackage[nolist]{acronym}
\usepackage{subcaption}
\usepackage{layouts}
\usepackage{multirow}
\usepackage{tikz}
\usetikzlibrary{positioning,arrows.meta,fit,calc,shapes.multipart}
\tikzset{
  layer/.style = {draw, thick, minimum height=1.1cm, minimum width=2.5cm, align=center, fill=#1!20},
  enc/.style   = {layer=blue},
  dec/.style   = {layer=green},
  bott/.style  = {layer=orange},
  io/.style    = {layer=gray},
  >={Stealth[length=6pt]},
  arrow/.style = {->, thick},
  tcnblock/.style = {
    draw,
    thick,
    minimum height=1.1cm,
    minimum width=2.6cm,
    align=center,
    rounded corners,
    fill=#1!20
  },
  conv/.style = {tcnblock=cyan},
  norm/.style = {tcnblock=violet},
  actv/.style = {tcnblock=lime},
  annotation/.style = {font=\footnotesize, align=center},
}

\setcopyright{acmlicensed}
\copyrightyear{2018}
\acmYear{2018}
\acmDOI{XXXXXXX.XXXXXXX}
\acmConference[Conference acronym 'XX]{Make sure to enter the correct
  conference title from your rights confirmation email}{June 03--05,
  2018}{Woodstock, NY}
\acmISBN{978-1-4503-XXXX-X/2018/06}

\newcommand{\bj}[1]{{\color{black} #1}}
\begin{document}

\title{A Real-Time Error Prevention System for Gaze-Based Interaction in Virtual Reality Based on Anomaly Detection}


\author{Björn R. Severitt}
\email{bjoern.severitt@uni-tuebingen.de}
\orcid{0009-0000-1343-4164}
\affiliation{%
	\institution{University of Tübingen}
	\streetaddress{Maria-von-Linden-Str 6}
	\city{Tübingen} 
	\country{Germany} 
	\postcode{72076}
}
\author{Yannick Sauer}
\email{yannick.sauer@zeiss.com}
\orcid{0000-0002-7513-341X}
\affiliation{%
	\institution{Carl Zeiss Vision International GmbH}
	\streetaddress{Turnstrasse 27}
	\city{Aalen} 
	\country{Germany} 
	\postcode{73430}
}
\author{Nora Castner}
\email{nora.castner@zeiss.com}
\orcid{0000-0002-6771-7693}
\affiliation{%
	\institution{Carl Zeiss Vision International GmbH}
	\streetaddress{Turnstrasse 27}
	\city{Aalen} 
	\country{Germany} 
	\postcode{73430}
}

\author{Siegfried Wahl}
\email{siegfried.wahl@uni-tuebingen.de}
\orcid{0000-0003-3437-6711}
\affiliation{%
	\institution{University of Tübingen}
	\streetaddress{Maria-von-Linden-Str 6}
	\city{Tübingen} 
	\country{Germany} 
	\postcode{72076}
}
\affiliation{%
	\institution{Carl Zeiss Vision International GmbH}
	\streetaddress{Turnstrasse 27}
	\city{Aalen} 
	\country{Germany} 
	\postcode{73430}
}

\renewcommand{\shortauthors}{Severitt et al.}


\begin{abstract}
    Gaze-based interaction enables intuitive, hands-free control in immersive environments, but remains susceptible to unintended inputs. We present a real-time error prevention system (EPS) that uses a temporal convolutional network autoencoder (TCNAE) to detect anomalies in gaze dynamics during selection tasks. In a visual search task in VR, 41 participants used three gaze-based methods—dwell time, gaze and head direction alignment, and nod—with and without EPS. The system reduced erroneous selections by up to 95\% for dwell time and gaze and head, and was positively received by most users. Performance varied for nodding and between individuals, suggesting the need for adaptive systems. Objective metrics and subjective evaluations show that anomaly-based error prevention can improve gaze interfaces without disrupting interaction. These findings demonstrate the potential of anomaly-based error prevention for gaze interfaces and suggest applications in VR, AR, and assistive technologies.
\end{abstract}



\received{20 February 2007}
\received[revised]{12 March 2009}
\received[accepted]{5 June 2009}

\maketitle

\section{Introduction}
Gaze-based interaction offers an intuitive and hands-free way to communicate with digital systems, especially in immersive environments and assistive technologies where traditional input devices are impractical \cite{severitt2024bidirectionalcommunication,sidenmark2021radi}. 
Despite its potential, gaze interaction still faces challenges in accurately interpreting the user's intentions. A well-known issue is the Midas touch problem \cite{jacob1990dwell}, in which unintended gaze behavior is mistakenly registered as intentional input.

To mitigate such errors, various confirmation mechanisms have been proposed, including head gestures, dwell times, and combined gaze-head alignment \cite{sidenmark2019eyehead,spakov2012enhanced,esteves2015orbits}. However, even with these mechanisms, false selections persist and can undermine user trust—especially in collaborative or safety-critical scenarios \cite{padovani2019inaccurate,flook2019errorimpact}.

Given previous findings that gaze behavior reflects signs of confusion or correction even before a selection is confirmed \cite{severitt2024communicationBreakdown}, we focus our anomaly detection approach on the immediate temporal vicinity of the selection event to enable early error detection. In our previous work \cite{severitt2025anomalydetection}, we presented a theoretical proof-of-concept for using anomaly detection to identify erroneous gaze-based selections, defining errors as statistical outliers in gaze dynamics. This concept was further substantiated by empirical results in \cite{severitt2025interplayuserpreferenceprecision}, where we showed that even with confirmation mechanisms, selection errors frequently occurred, negatively impacting the user experience.

At the same time, the success of gaze-based interaction depends not only on technical accuracy, but also on the quality of the user experience. Error prevention mechanisms may be most effective when they work in the background, reducing errors without disrupting interaction. Ideally, users can continue to focus on their task while the system silently supports them, contributing to a more natural and trustworthy experience. From this perspective, both reliable anomaly detection and unobtrusive support are important. The goal of this work is therefore not only to investigate the accuracy of error prevention, but also to explore how such systems can contribute to smoother eye-based interaction, with potential implications for consumer and assistive technologies.


On this basis, the present study evaluates a real-time \ac{eps} based on an \ac{ae} with a \ac{tcn}, a model architecture that has proven effective for anomaly detection in time series data \cite{he2019tcnanomaly}. The system is trained to reconstruct typical gaze patterns for correct selections, with deviations from this learned representation interpreted as potential errors. We implement and test the \ac{eps} in a \ac{vr} game using a visual search task and compare three gaze-based interaction methods—Dwell Time, Gaze and Head and Nod—with and without the system. Specifically, we address two central questions: (1) How do participants experience and evaluate the system in terms of usefulness, trust, and overall preference, and (2) can the \ac{eps} reliably prevent errors in a real-time environment across different interaction methods?

\section{Related Work}



Gaze-based interaction has become increasingly important in various fields, including assistive technologies \cite{parisay2021eyetap, istance2008snapclutch, rajanna2016gawschi}, \acf{ar}/\acf{vr} \cite{mohan2018dualgaze, velloso2016ambigaze, gardony2024characterizing, schweigert2019eyepointing}, and \ac{hci} in general \cite{gemicioglu2023gazetongue}. An important goal of these approaches is to reduce unintended selections, known as the \textit{Midas touch problem}, in order to improve both the user experience and trust in the system. Recent studies also show that features derived from eye-tracking data can be used to infer user intentions \cite {david2021towards, bovo2020detecting} and detect interaction errors \cite{severitt2023leveraginggaze, bovo2020detecting, lufthansa}, highlighting the potential of gaze as a powerful input modality.

Recent studies have demonstrated the potential of machine learning to improve gaze-based input by contextualizing gaze behavior within the interaction environment. For example, \citet{shevtsova2025gazebasedinteractionml} shows that machine learning techniques were applied to gaze data with free behavior in a visually complex game environment, resulting in a significant reduction in false alarms and an improvement in user satisfaction—all without explicit intention confirmation. Such results underscore the increasing suitability of data-driven methods for improving gaze-based interaction.

In addition, sequence modeling techniques such as \ac{tcn} were used to classify user errors and predict intentions directly from raw gaze histories, achieving performance well above chance levels \cite{sendhilnathan2022detectininputrecognitionerrors}. Continuing this work, \citet{sendhilnathan2025multimodal} demonstrated that incorporating multimodal inputs into the \ac{tcn} framework further improves error classification. Although these approaches are effective, they are generally based on a fixed taxonomy of error classes, which makes it difficult to generalize them to previously unknown or unexpected interaction errors.

In contrast, our approach reframes selection errors not as predefined classes, but as anomalies within the continuous temporal gaze signal. By treating errors as rare, statistically distinct events, we align our methodology with established techniques in time series anomaly detection, as surveyed by \citet{schmidl2022anomalydetectionsummary}. In our previous work, we demonstrated a theoretical proof-of-concept for applying anomaly detection algorithms to gaze-based interaction data \cite{severitt2025anomalydetection}. In this study, we extend that foundation by implementing and evaluating anomaly detection models on real gaze recordings from a target selection task in a \ac{vr} environment. 

\section{Methods}

In this section, we explain the anomaly detection method we use for the \ac{eps} and the \ac{ve} to test this method, including the various gaze-based interaction methods.

\subsection{Anomaly detection model}

In order to reliably distinguish between correct and incorrect selections, we use an anomaly detection approach that models typical gaze behavior for correct selections. For this purpose, we use a \ac{tcnae} that captures  dependencies in gaze dynamics and provides a reconstruction-based measurement of selection validity. In the following subsection, we describe the model architecture, input representation, and training procedure in detail.

\begin{figure*}
    \centering
    \begin{tikzpicture}[node distance=1cm and 1cm]

    \node[io] (input) {Input\\$\mathbb{R}^{36}$};
    \node[enc, below=of input] (enc1) {TCNBlock\\$1 \rightarrow 4$\\$k{=}5,\;d{=}1$};
    \node[enc, below=of enc1] (enc2) {TCNBlock\\$4 \rightarrow 8$\\$k{=}5,\;d{=}2$};
    \node[enc, below=of enc2] (flat)   {Flatten\\$8{\times}36 \rightarrow 288$};
    
    
    \node[bott, below=of flat] (fcenc)  {FC: $288 \rightarrow 10$};
    \node[bott, right=of fcenc] (fcdec) {FC: $10 \rightarrow 288$};
    
    \node[dec, above=of fcdec] (reshape) {Reshape\\$288 \rightarrow 8{\times}36$};
    \node[dec, above=of reshape]  (dec1) {TCNBlock\\$8 \rightarrow 4$\\$k{=}5,\;d{=}1$};
    \node[dec, above=of dec1]     (dec2) {TCNBlock\\$4 \rightarrow 1$\\$k{=}5,\;d{=}2$};
    \node[io, above=of dec2]      (output) {Output\\$\mathbb{R}^{36}$};
    
    \draw[arrow] (input) -- (enc1);
    \draw[arrow] (enc1) -- (enc2);
    \draw[arrow] (enc2) -- (flat);
    \draw[arrow] (flat) -- (fcenc);
    \draw[arrow] (fcenc) -- (fcdec);
    
    \draw[arrow] (fcdec) -- (reshape);
    \draw[arrow] (reshape) -- (dec1);
    \draw[arrow] (dec1) -- (dec2);
    \draw[arrow] (dec2) -- (output);
    
    \node[draw, thick, rounded corners, fit=(enc1)(flat), label={[rotate=90, yshift=1.3em, anchor=center]left:\textbf{Encoder}}] {};
    \node[draw, thick, rounded corners, fit=(reshape)(dec2), label={[rotate=90, yshift=-1.3em, anchor=center]right:\textbf{Decoder}}] {};
    \node[draw, thick, rounded corners, fit=(fcenc)(fcdec), label=below:\textbf{Bottleneck}] {};

    \node[conv, right=2cm of dec2] (conv) {Conv1D\\$ch_{in}\rightarrow ch_{out}$\\$k=\text{kernel size}$\\$\ d=\text{dialation}$};
    \node[norm, below=of conv] (bn) {BatchNorm1D};
    \node[actv, below=of bn] (relu) {ReLU};
    
    \draw[arrow] (conv) -- (bn);
    \draw[arrow] (bn) -- (relu);

    \node[draw, thick, rounded corners, fit=(conv)(relu), label=above:\textbf{TCNBlock}]{};
    
    \end{tikzpicture}
    \caption{Architecture of the \ac{tcnae}.
The network comprises an encoder (left), bottleneck (center), and decoder (right). The encoder and decoder each consist of stacked TCN blocks, where each block includes a 1D convolution, batch normalization, and \ac{relu} activation (see TCNBlock, right). The number of input and output channels, kernel size, and dilation factor are indicated within each block. The bottleneck compresses the input sequence into a fixed-length latent vector, which is then reconstructed to the original shape. All convolutional layers preserve sequence length via appropriate padding.}
    \label{fig:TCNAEArchitecture}
\end{figure*}
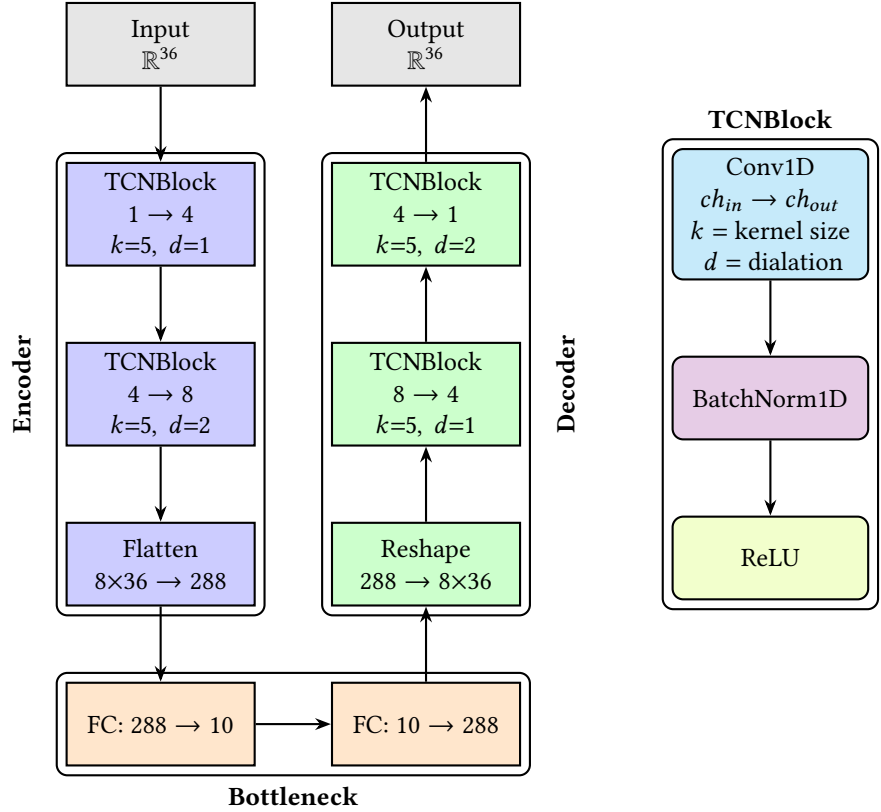

\paragraph{Method} The anomaly detection method we used to find anomalies in gaze behavior shortly after selection is an \ac{ae} based on a \ac{tcn} implemented with PyTorch~\cite{pytorch2019paszke}. The architecture is shown in Fig.~\ref{fig:TCNAEArchitecture}. The idea is to train the network to reconstruct the input of normal selections. Non-normal selections—in this case, anomalies—are therefore never trained for reconstruction, which should result in a higher reconstruction error \cite{schmidl2022anomalydetectionsummary}. To determine whether a selection was intentional, we computed the reconstruction error $err \in \mathbb{R}$ between the input $in \in \mathbb{R}^{36}$ and the corresponding output of the network $out \in \mathbb{R}^{36}$. Based on this error, the selection validity $s \in \{\text{correct},\text{incorrect}\}$ was determined as follows:
\begin{align}
    s = 
    \begin{cases}
        \text{correct} & \text{if} \quad err < th \\
        \text{incorrect} & \text{if} \quad err \geq th
    \end{cases}
\end{align}
Here, $th \in \mathbb{R}$ denotes a fixed threshold for the reconstruction error.

\begin{figure}
    \centering
    \includegraphics[width=\linewidth]{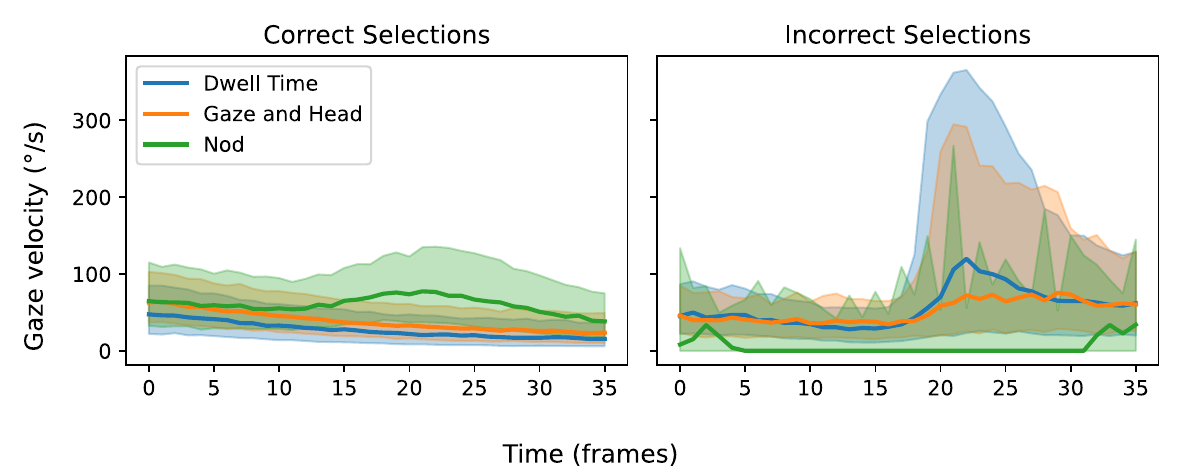}
    \caption{Velocity profiles of network input for the various interaction methods, represented as the median value of all participants at each point in time, with the shaded areas indicating the 25th and 75th percentiles.}
    \label{fig:velocityProfiles}
\end{figure}

\paragraph{Input} In our previous work, in which we created a proof of concept based on our existing data, we used a time series of vectors between consecutive gaze vectors collected shortly before and after selection as input \cite{severitt2025anomalydetection}. This approach is problematic when frame losses occur, as the angle may then be much larger, even if the gaze behavior does not change. For this reason, we changed the input to angular velocity. In practice, we used the last 37 gaze vectors, denoted as $gv_i \in \mathbb{R}^3$ with $i= 0, \dots, 36$, which were recorded 200 ms after the selection event. Due to the frame rate of 90~fps, this window covers gaze data approximately 200~ms both before and after the selection event, assuming no frames were dropped. Using these vectors, we computed a time series of angular velocities $v_j \in \mathbb{R}$ for $j = 0,\dots, 35$, defined as:
\begin{align*}
    a_j &= \frac{180}{\pi} \arccos(gv_j \cdot gv_{j+1}) \quad \text{with} \quad \|gv_i\| =1\quad \forall i=0,\dots,36\\
    v_j &= \frac{a_j}{(t_{j+1} - t_j) \cdot 10^{-3}}
\end{align*}
Here, $a_j$ represents the angular distance in degrees between consecutive normalized gaze vectors, and $t_j$ denotes the timestamp of $gv_j$ in milliseconds. The angular velocity $v_j$ is expressed in degrees per second, with the resulting profiles shown in Fig.~\ref{fig:velocityProfiles}.


\paragraph{Training} 
To train the network, we use data from our previous experiment \cite{severitt2025interplayuserpreferenceprecision}, which uses the same environment, which is explained in more detail below (see subsection~\ref{subsec:game}). We captured all correct selection events from the 52 participants and collected the necessary gaze vectors and timestamps surrounding the selection event. With this data, we trained the network to reconstruct correct selections as described above. To find the hyperparameters, we performed five-fold cross-validation, following the procedure of our proof-of-concept study \cite{severitt2025anomalydetection}. Cross-validation was performed separately for each interaction method, and the results were then aggregated across methods to determine the hyperparameter configuration that achieved the best overall performance.
We used \ac{mse} as the loss function, which will also be used later as the reconstruction error. We trained the network for 2000 epochs with a batch size of 4000, a learning rate of $10^{-2}$, and using the Adam optimizer. We trained a separate network for each gaze-based interaction method and determined an individual reconstruction error threshold for each. All other hyperparameters were kept identical across networks. The threshold was computed by evaluating the \ac{mse} for all correct selections in the training dataset and selecting the 95th percentile of these values.


\subsection{Game} \label{subsec:game}
To evaluate the \ac{eps} for each interaction method, we integrated it into the \ac{vr} game environment developed in our previous study \cite{severitt2025interplayuserpreferenceprecision}. We only detail briefly aspects of the game that are relevant to understanding the rest of the paper, though, for more detail, see the previously mentioned study. Participants were instructed to achieve the highest possible score. 

\begin{figure}
    \centering
    \begin{subfigure}[t]{0.45\linewidth}
        \centering
        \includegraphics[width=\linewidth]{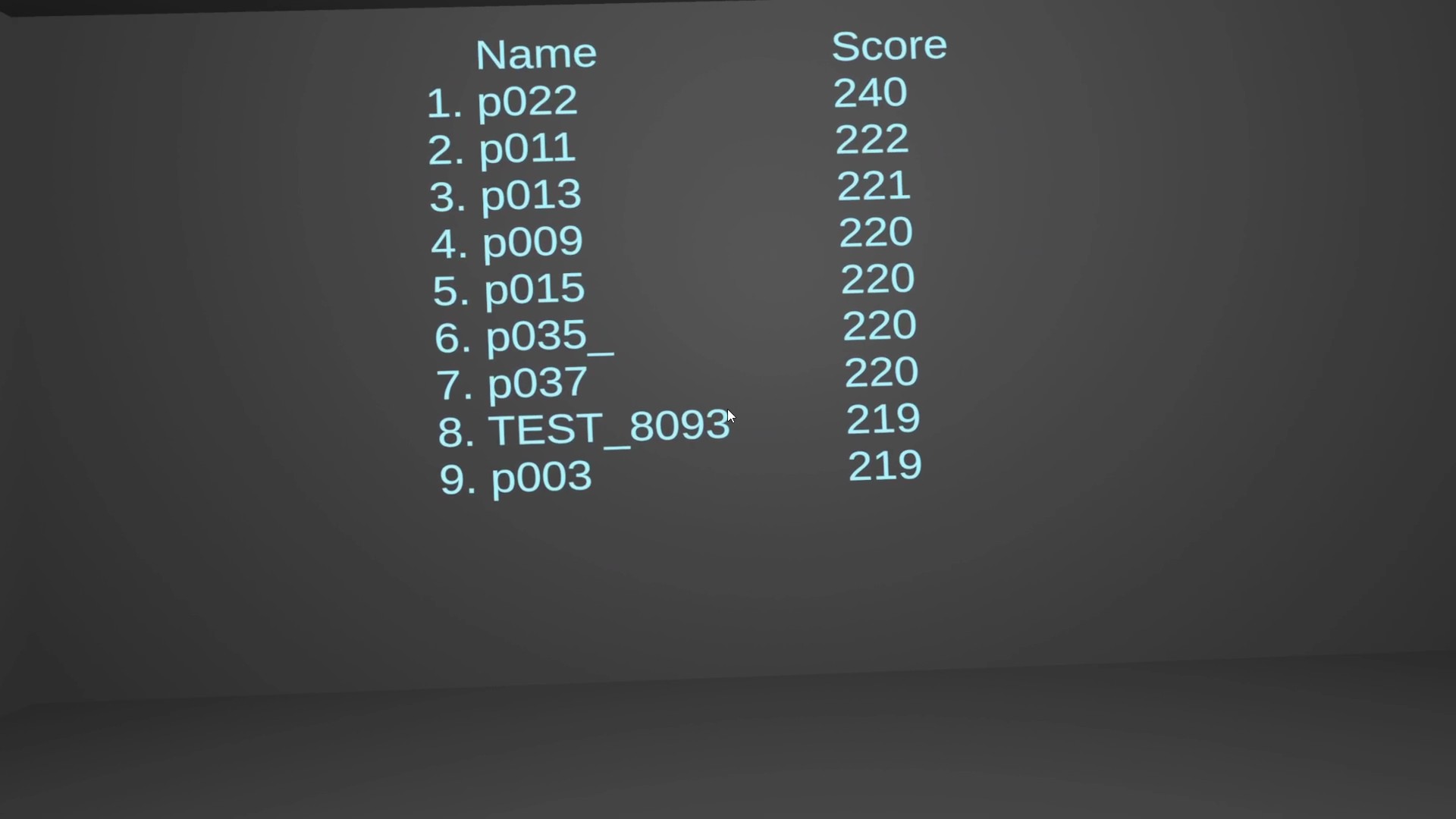}
        \caption{Highscorelist}
        \label{subfig:highscorelist}
    \end{subfigure}
    \hfill
    \begin{subfigure}[t]{0.45\linewidth}
        \centering
        \includegraphics[width=\linewidth]{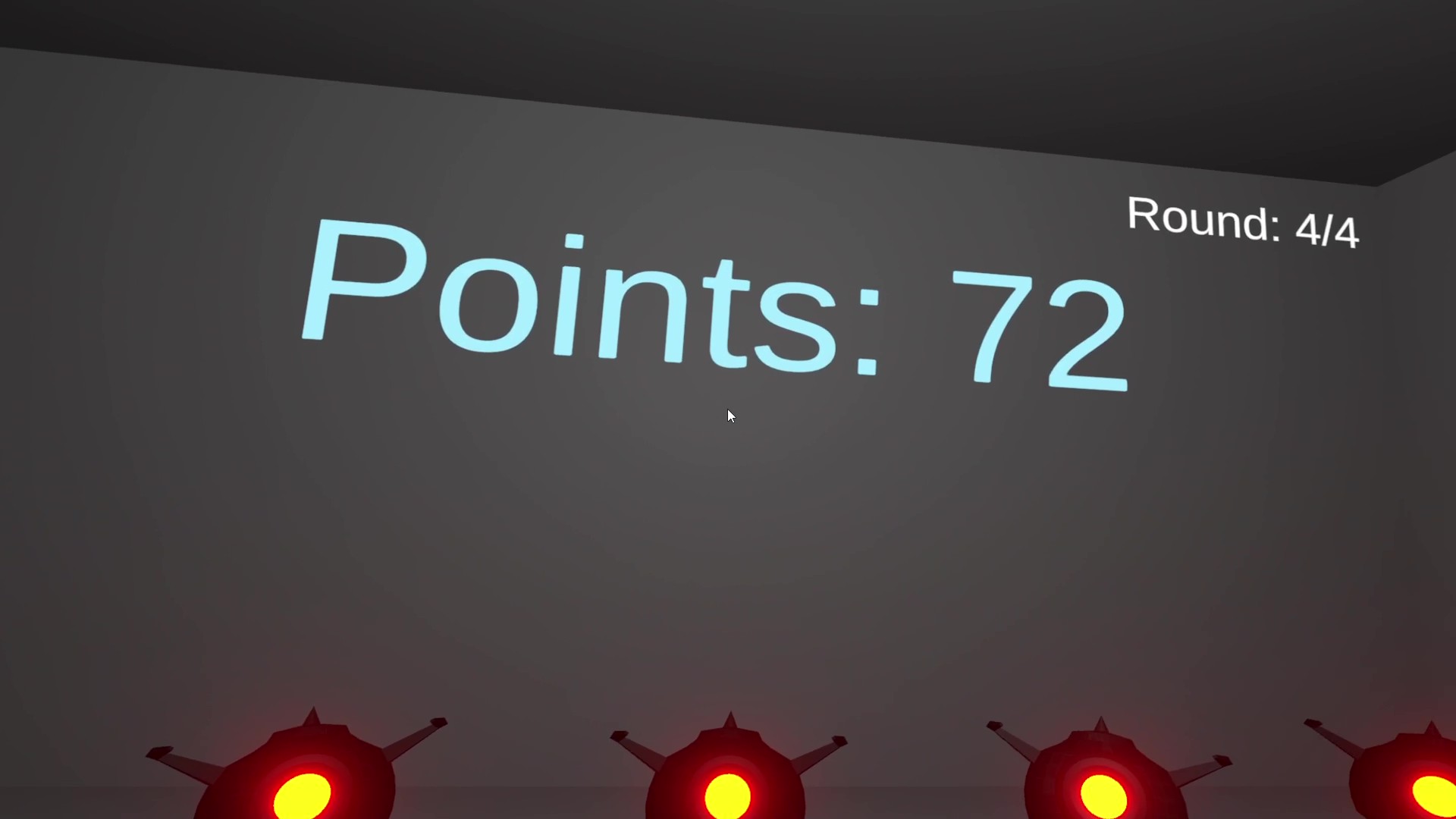} 
        \caption{Current points}
        \label{subfig:highscorelist}
    \end{subfigure}
    \hfill
    \begin{subfigure}[t]{0.45\linewidth}
        \centering
        \includegraphics[width=\linewidth]{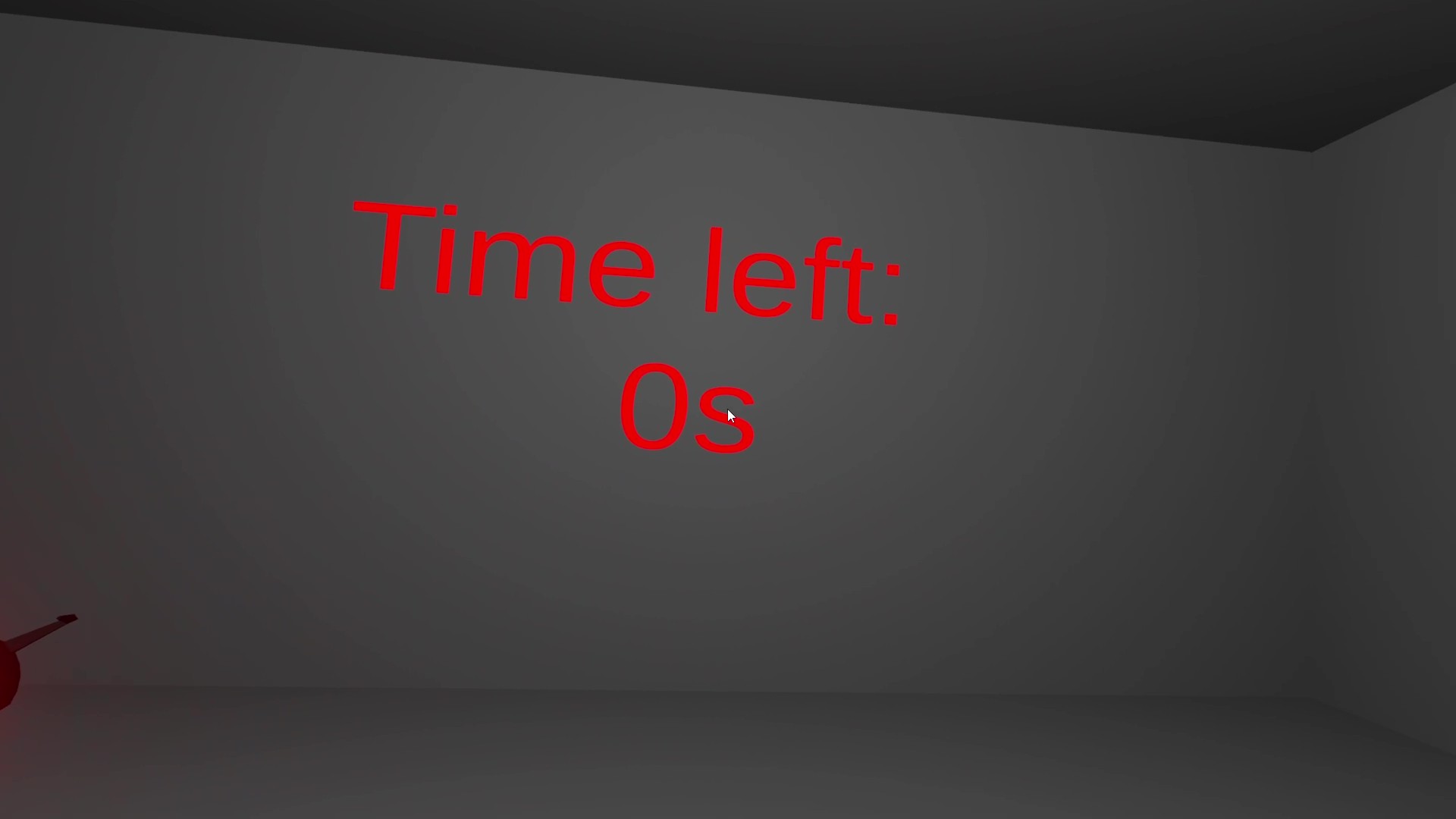}
        \caption{Timer}
        \label{subfig:highscorelist}
    \end{subfigure}
    \hfill
    \begin{subfigure}[t]{0.45\linewidth}
        \centering
        \includegraphics[width=\linewidth]{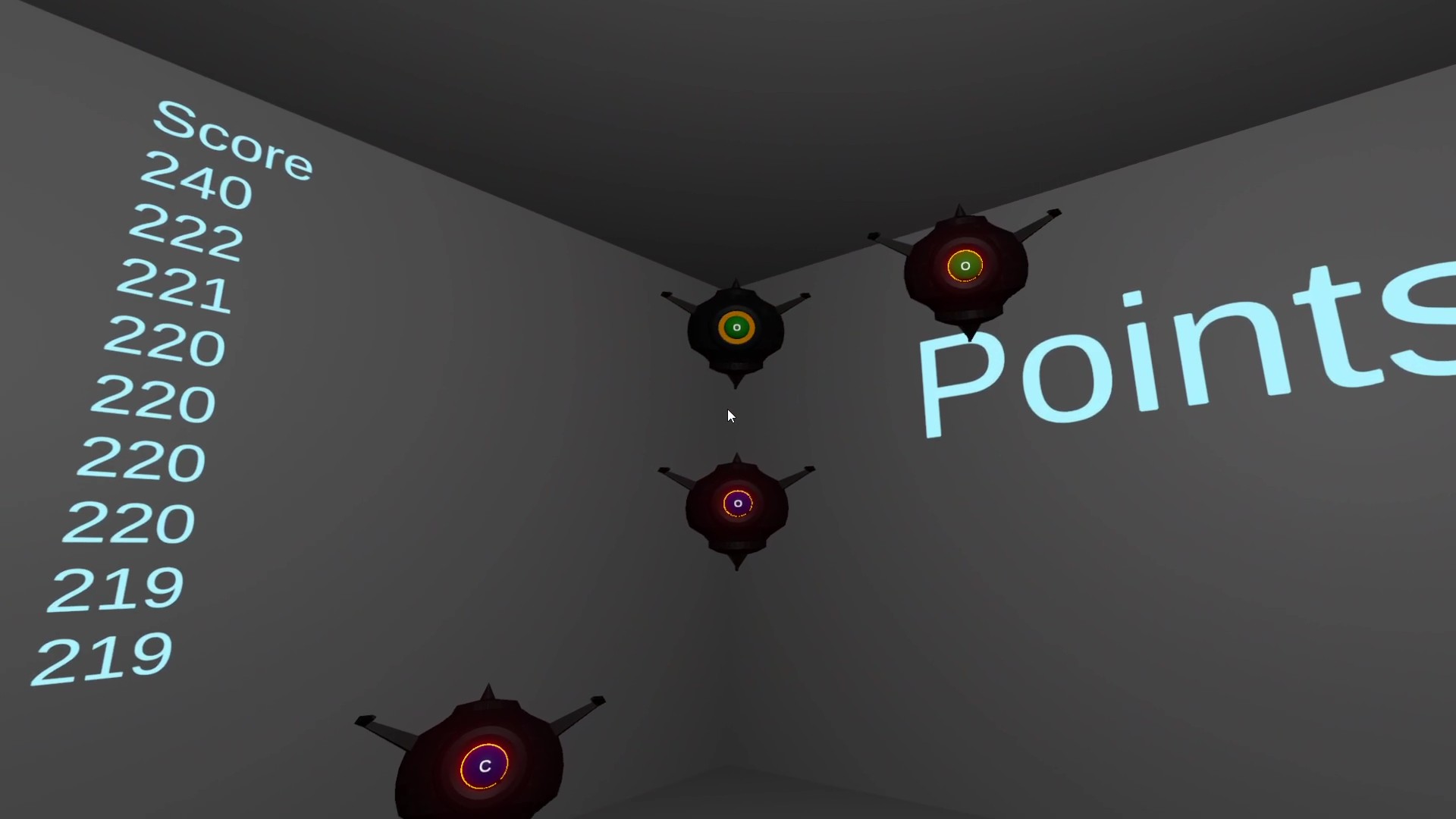}
        \caption{Robots with targets}
        \label{subfig:highscorelist}
    \end{subfigure}
    \caption{Illustration of \ac{ve}: (a) high score list, (b) current points and lap counter (top right), (c) timer display, and (d) robots with targets, with the main target marked by the “C” at the bottom.}
    \label{fig:ve}
\end{figure}

\paragraph{Game environment}
The \ac{ve} consisted of a square-shaped room with interactive information displayed on the walls. One wall featured a high score leaderboard to motivate participants. Another wall displayed the current score in the center, along with a round counter positioned in the top right corner. A third wall showed a countdown timer for the active round. The overall layout was designed to provide continuous feedback and maintain user engagement, and is shown in Fig.~\ref{fig:ve}.

\paragraph{Gameplay}
The core mechanics of the game involve identifying a target among flying robots positioned in the virtual environment. At the beginning of each round, robots move to one of approximately 18 predefined positions arranged within a 180-degree field in front of the user. 
Once a robot reaches its position, a sphere appears in front of it, displaying a character: the target is indicated by the letter “C”, while distractors are indicated by the letter “O”. Participants must identify and select the correct target using the gaze-based selection method assigned to the current experimental block. Selecting the correct target incurs no penalty, whereas incorrectly selecting a distractor results in a deduction from the score.
After a correct selection, the robots reposition themselves by flying to a new position at the predefined locations, and the process repeats itself. Each round lasts 30 seconds, during which participants attempt to identify as many targets as possible. The experiment consists of three blocks, each corresponding to a different gaze selection method.

\paragraph{Scoring System.}
Participants received points based on their performance according to the following rules. Successfully destroying the correct target awarded between 5 and 20 points, depending on the response time. If the target was destroyed after more than 10 seconds, the minimum of 5 points was awarded. For faster responses, the score was computed using linear interpolation, with shorter reaction times yielding higher point values up to a maximum of 20.
Incorrect selections — specifically destroying a distractor — resulted in a penalty of 21 points. This scoring scheme was designed to encourage both accuracy and speed, introducing time pressure to promote focused and efficient behavior.

\paragraph{Gaze-based selection Methods.}
Three gaze-based selection methods were compared in the study. The first method, \textit{Dwell time}, allows participants to select a target by fixating on it for a specified duration, as originally introduced by \citet{jacob1990dwell}. The second method, \textit{Gaze and Head}, requires both the gaze direction and head orientation to remain aligned with the target for the same duration, following the approach proposed by \citet{sidenmark2019eyehead}. The third method, \textit{Nod}, involves a gaze fixation followed by a nod gesture to confirm the selection, based on the work of \citet{spakov2012enhanced}.  The dwell time for the Dwell Time method was set to 0.4 seconds and for Gaze and Head to 0.3 seconds.
Since the anomaly detection models were trained independently for each selection method, we refer to these three input techniques as distinct conditions in the context of the anomaly detection results.

\subsection{Study}

In this subsection, we provide details on the study design, including participant demographics, \ac{vr} hardware and software environment, and overall experimental setup. 

\begin{table*}[]
    \begin{tabular}{ll|rrr}
    \toprule
     &  & General & Women & Men \\
    Topic & Category &  &  &  \\
    \midrule
    Count & Count & 37 & 26 & 11 \\
    \cline{1-5}
    \multirow[c]{4}{*}{Age} & Mean age & 26.14 & 26 & 26.45 \\
     & Std age & 4.74 & 4.05 & 6.07 \\
     & Minimal age & 19 & 20 & 19 \\
     & Maximal age & 39 & 33 & 39 \\
    \cline{1-5}
    \multirow[c]{3}{*}{Visual Aids} & Contact lenses & 3 & 3 & 0 \\
     & Glasses & 13 & 7 & 6 \\
     & No & 21 & 16 & 5 \\
    \cline{1-5}
    \multirow[c]{5}{*}{VR Experience} & Expert & 0 & 0 & 0 \\
     & A lot & 4 & 1 & 3 \\
     & Moderate & 5 & 3 & 2 \\
     & Little & 17 & 13 & 4 \\
     & Not at all & 11 & 9 & 2 \\
    \cline{1-5}
    \multirow[c]{5}{*}{ETK Experience} & Expert & 0 & 0 & 0 \\
     & A lot & 1 & 1 & 0 \\
     & Moderate & 4 & 1 & 3 \\
     & Little & 21 & 15 & 6 \\
     & Not at all & 11 & 9 & 2 \\
    \bottomrule
    \end{tabular}
    \caption{Participant demographics and self-reported experience. The table summarizes age, use of visual aids, and prior experience with \acf{vr} and \acf{et}, separated by gender.}
    \label{tab:PatMetaData}
\end{table*}
\paragraph{Participants} A total of 41 participants from the University of Tübingen and the surrounding area were recruited for this study; four participants dropped out due to technical issues. The remaining participants had a mean age of $26.71 \pm 5.15$ years.
Of the remaining participants, 26 were women (mean age $26 \pm 4.05$ years) and 11 were men (mean age $26.45 \pm 6.07$ years). Most participants reported little or no prior experience with \ac{et} or \ac{vr}. Further details on participant characteristics are provided in Table~\ref{tab:PatMetaData}. 
This study was reviewed and approved by Faculty of Medicine at the University of Tübingen with a corresponding ethical approval identification code 247/202BO2. Participants provided their written informed consent to participate in this study.

\paragraph{\ac{vr} setup} 
The experiment was done with an HTC Vive Pro Eye head-mounted display (HMD; HTC Corporation, Taoyuan, Taiwan), which features a built-in Tobii eye tracker (Core SW 2.16.4.67) with an estimated accuracy of $0.5^\circ$ to $1.1^\circ$ and a sampling frequency of 120 Hz. Eye tracking data was calibrated and accessed using the Vive SRanipal SDK (HTC Corporation, Taoyuan, Taiwan). To extract gaze data from the \ac{vr} headset, we employed the ZERO-Interface~\cite{zero2023hosp}, which is integrated into the VisionaryVR~\cite{hosp2024visionaryvr} framework. This interface provides separate three-dimensional gaze vectors for each eye, as well as a combined gaze vector. The data were accessible in real time and used for both gameplay and executing the \ac{eps}. Additionally, all gaze data were recorded for post-hoc analysis.
An HTC Vive Pro Controller 2.0 was used for handheld interactions, such as adjusting sliders to complete questionnaires.

\paragraph{Design} 
At the beginning of each condition, participants were presented with a virtual text panel inside the \ac{vr} environment, explaining how the respective gaze-based interaction method worked. This was followed by a training round using the method without the \ac{eps}, allowing participants to become familiar with the interaction technique.
After the training round, another virtual instruction panel appeared, informing the participant that the next round would include the \ac{eps}. Upon completing this second round, a new panel prompted the participant with the question: “In which round was the error prevention system used?” Participants were required to select either First or Second. 
Following the training phase, the main experiment began. Participants completed five blocks, each consisting of two rounds—one with and one without the \ac{eps}. The order of the conditions within each block was randomized. After each block, participants again answered the same question to indicate which round they believed included the \ac{eps}. This served as a manipulation check to assess whether the system was noticeable.
After completing the five blocks for a given interaction method, participants were transported to a new virtual scene, where they completed a questionnaire using \ac{vr}-based \ac{ui} elements. This entire procedure was repeated for all three gaze-based interaction methods, with the method order randomized across participants to counterbalance potential learning effects.

\subsection{Measurements}

In order to comprehensively evaluate the effects of the \ac{eps}, we collected both subjective and objective data. Subjective measurements focused on participants' perceptions and experiences, while objective measurements captured task performance and system accuracy. 

\paragraph{Subjective measurements} We collected data using a questionnaire that was completed after each method. The questionnaire evaluated six aspects of the \ac{eps}: usefulness (reduction of accidental selections), confidence (deletion of incorrect selections only), frustration (annoyance caused by the system), confidence boost (increased confidence when using gaze-based interaction), interference (impairment of intended actions), and preference (willingness to use the system in future gaze-based interfaces). All items were rated on a 10-point scale bounded by opposite statements. The complete questionnaire can be found in the appendix. At the end of the experiment, participants additionally indicated their preferred method (Gaze Dwell, Gaze and Head, or Nod) and whether they found the \ac{eps} helpful (yes, no, not sure). 

\paragraph{Objective measurements} This was done in two ways. First, the performance of the participants was evaluated by comparing the conditions with and without the \ac{eps} for each method. Performance was evaluated based on the total number of points achieved and the number of incorrect selections. Next, the performance of the \ac{eps} itself was analyzed. This evaluation took into account classification accuracy, defined as the proportion of correct selections that were classified as correct and the proportion of incorrect selections that were classified as incorrect, reported separately for each method. 

\subsection{Data analysis}
Given the non-normal distribution of the data and the within-subjects design, non-parametric methods were applied. For comparisons across multiple related groups, the Friedman test was used as a non-parametric alternative to repeated-measures ANOVA \cite{anova1937friedman}.  For pairwise comparisons, the Wilcoxon signed-rank test was employed, which provides a non-parametric equivalent to the paired-samples t-test \cite{comparison1945wilcoxon}. To control for Type I error inflation in cases of multiple comparisons, the Bonferroni correction was applied \cite{multiple1961dunn}. \bj{Eta-squared ($\eta^2$) was used as a measure of effect size in the Friedman test; for pairwise Wilcoxon comparisons, the effect size $r$ was calculated based on a z-transformed approximation of the p-value \cite{cohen2013statistical}}. All statistical analyses were conducted in Python (version 3.13.5) using NumPy~\cite{numpy2020harris}, pandas~\cite{pandas2010mckinney}, SciPy~\cite{scipy2020virtanen}, and statsmodels~\cite{statsmodels2010seabold}. In addition, matplotlib~\cite{matplotlib2007hunter} and seaborn~\cite{seaborn2021waskom} were used for data visualization.

\section{Results}

Below, we present the results of our study, in which we combine subjective evaluations with objective performance measurements. The subjective results reflect participants' perceptions of the various interaction methods and the \ac{eps}, while the objective results provide quantitative evidence of task performance and system accuracy.

\subsection{Subjective measurements}

\begin{table*}[]
    \centering
    \begin{subtable}[t]{0.45\linewidth}
        \centering
        \begin{tabular}{lrrr}
        \toprule
         & Dwell Time & Gaze and Head & Nod \\
        \midrule
        Women & 7 & 11 & 8 \\
        Men & 4 & 5 & 2 \\
        General & 11 & 16 & 10 \\
        \bottomrule
        \end{tabular}
        \caption{Preference}
        \label{subtab:PrefMethod}
    \end{subtable}
    \hfill
    \begin{subtable}[t]{0.45\linewidth}
        \centering
        \begin{tabular}{lrrr}
        \toprule
         & Yes & No & Not sure \\
        \midrule
        Women & 20 & 1 & 5 \\
        Men & 11 & 0 & 0 \\
        General & 31 & 1 & 5 \\
        \bottomrule
        \end{tabular}
        \caption{Helpfulness}
        \label{subtab:EPHelpful}
    \end{subtable}
    \caption{Results of the post-experiment questionnaire. Sub-table (a) presents participants’ preferred interaction method, separated by gender and overall counts. Sub-table (b) summarizes participants' responses to whether they perceived the \ac{eps} as helpful.}
    \label{tab:PostExperimentQuestions}
\end{table*}

The distribution of participants' preferred interaction methods is shown in Table~\ref{subtab:PrefMethod}. In general, Gaze and Head was chosen most frequently, followed by Nod and Dwell Time. The evaluation of the \ac{eps} is summarized in Table~\ref{subtab:EPHelpful}, where the majority of participants reported that the system was helpful, while only a small number expressed uncertainty, and just a single participant rated it as not helpful. 

\begin{figure*}
    \centering
    \includegraphics[width=1\linewidth]{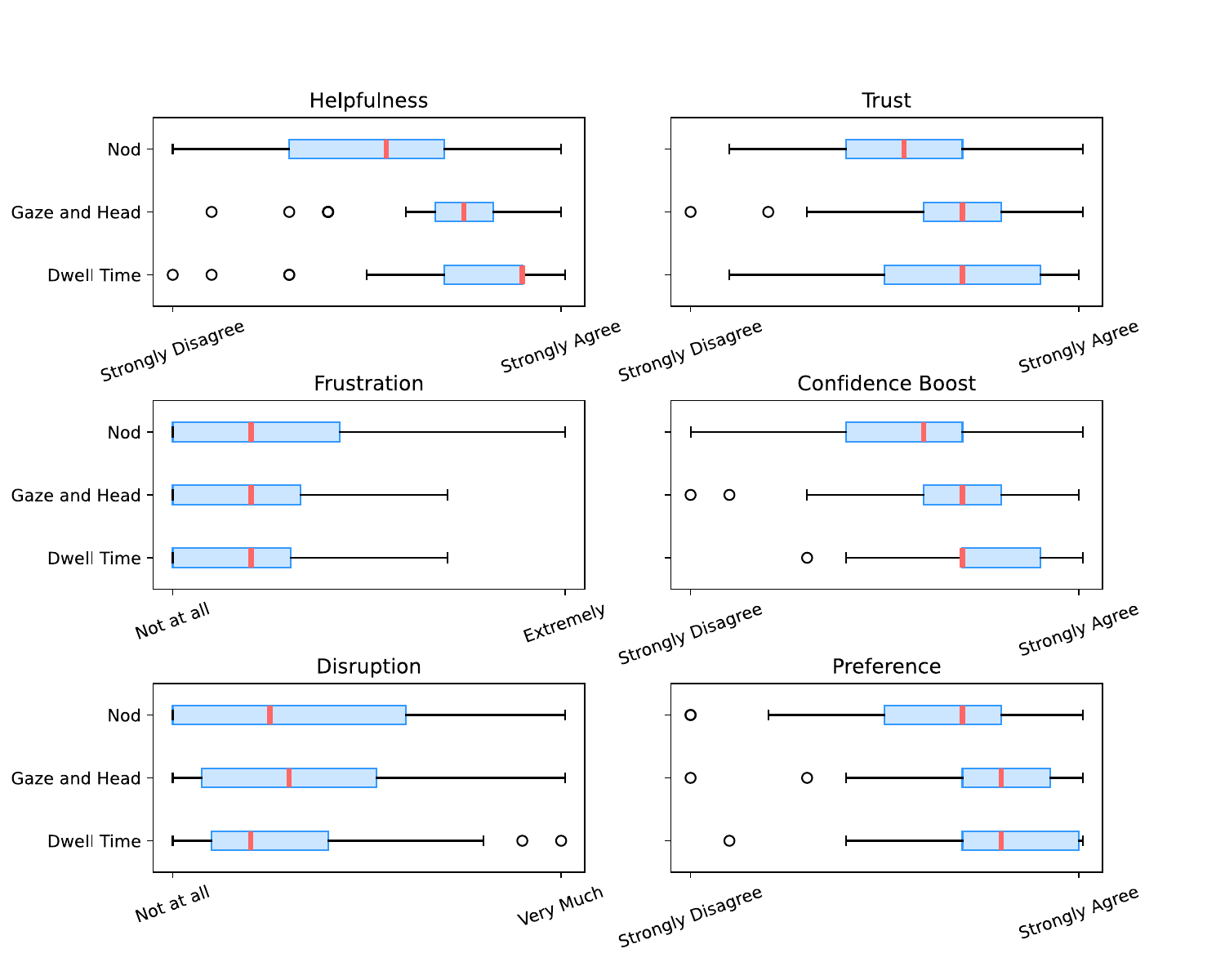}
    \caption{Boxplots of participants' responses to the questionnaire for each method. Each subplot corresponds to a questionnaire item, with the methods shown on the y-axis and the response scale on the x-axis. The figure illustrates the distribution, central tendency, and variability of all participants' ratings.}
    \label{fig:QuestsAnswers}
\end{figure*} 

The participants' ability to recognize whether \ac{eps} was active varied depending on the method: accuracy was $79.5\%$ for Dwell Time, $70.8\%$ for Gaze and Head, and $55\%$ for Nod, which is only slightly better than chance.
In Fig.~\ref{fig:QuestsAnswers}, the responses to the questionnaire for the different methods are shown. The analysis revealed significant differences between methods for helpfulness ($\chi^2 = 20.21, p < 0.001, \eta^2=0.289$), trust ($\chi^2=8.92, p=0.012, \eta^2=0.127$), confidence boost ($\chi^2=9.88, p=0.007, \eta^2=0.141$), and preference ($\chi^2=14.75, p=0.001, \eta^2=0.211$). Post-hoc comparisons showed that the Nod method received significantly lower ratings than Dwell Time and Head and Gaze across these dimensions. Descriptively, Dwell Time generally had the highest median and mean ratings for helpfulness, confidence boost, and preference, while Nod had the lowest, and Head and Gaze fell in between. No significant differences were found for frustration ($\chi^2=0.16, p=0.925, \eta^2=0.008$), or disruption ($\chi^2 = 0.23, p=0.893, \eta^2=0.007$), where median ratings were similar and generally low for frustration and disruption. 

\subsection{Objective measurements}

\begin{figure*}
    \centering
    \begin{subfigure}[t]{\linewidth}
        \centering
        \includegraphics[width=\linewidth]{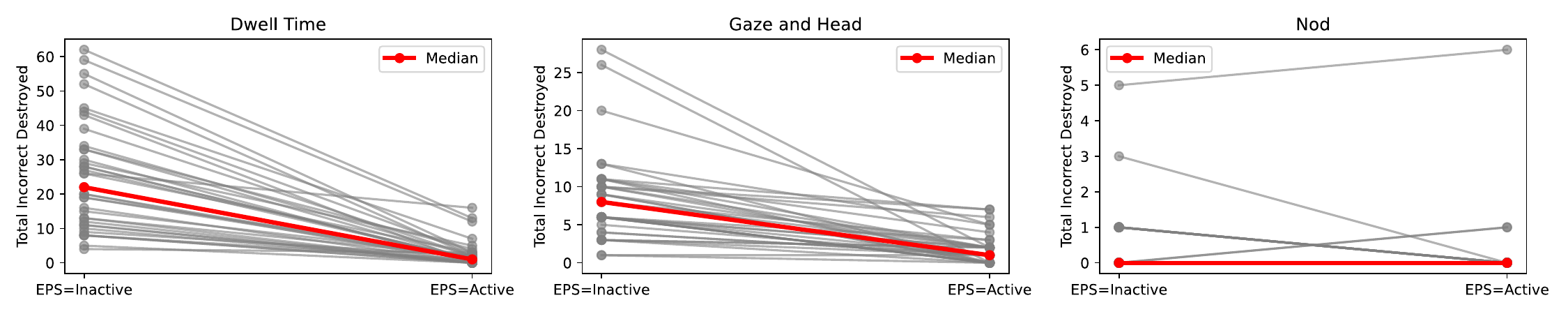}
        \caption{}
        \label{subfig:PairedSelections}
    \end{subfigure}
    \begin{subfigure}[t]{\linewidth}
        \centering
        \includegraphics[width=\linewidth]{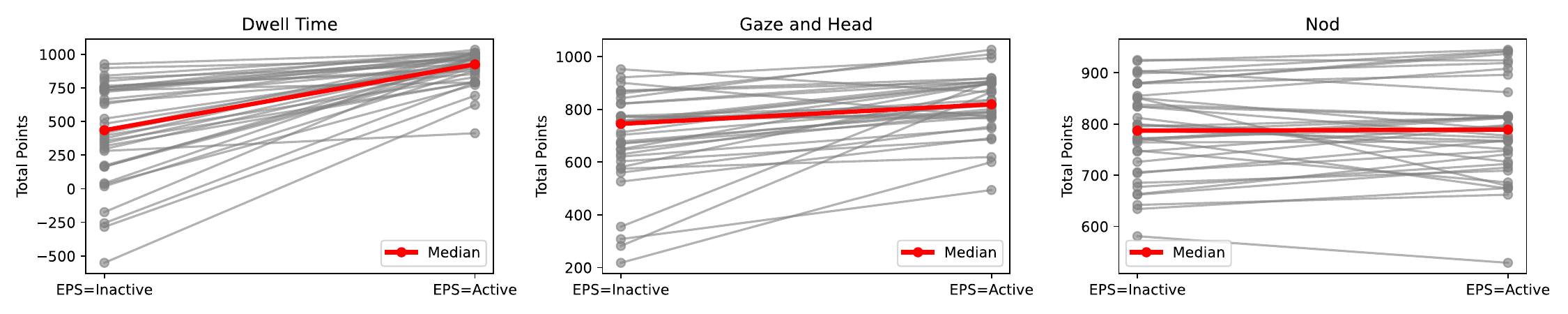}
        \caption{}
        \label{subfig:PairedPoints}
    \end{subfigure}
    \caption{Comparison of task performance with the \ac{eps} inactive and active. Subfigure (a) shows the number of incorrect selections, while subfigure (b) displays the median achieved points.}
    \label{fig:PtPerformance}
\end{figure*}

The measurements of participant performance are shown in Fig.~\ref{fig:PtPerformance}. For the number of incorrect selections (summed over all rounds), significant differences were found across all three methods (see Fig.~\ref{subfig:PairedSelections}). With the \ac{eps} active, participants made markedly fewer incorrect selections. In the Dwell Time method, the median dropped from 22 without \ac{eps}  to 1 with \ac{eps}  ($W = 0.0, p < 0.001, r=0.872$). A similar effect was observed for Head and Gaze, with medians of 8 and 1, respectively ($W = 0.0, p < 0.001, r=0.861$). For Nod, the Wilcoxon signed-rank test also indicated a significant reduction in errors ($W = 22.5, p = 0.009, r=0.387$), although the median remained 0 in both conditions.
A comparable pattern was found for the points achieved (see Fig.~\ref{subfig:PairedPoints}). Participants gained substantially more points when the \ac{eps}  was active. For Dwell Time, the median increased from 435 to 925 ($W = 703.0, p < 0.001, r=0.872$), and for Head and Gaze, from 745 to 819 ($W = 667.0, p < 0.001, r=0.783$). In contrast, no significant difference was observed for Nod, where the medians were nearly identical at 787 without \ac{eps} and 789 with \ac{eps} ($W = 409.0, p = 0.196, r= 0.140$).

\begin{table*}[]
    \centering
    \begin{tabular}{lrrrr}
    \toprule
     & Correct accuracy & Incorrect accuracy & Macro & Micro \\
    \midrule
    Dwell Time & 0.794 & 0.928 & 0.861 & 0.829 \\
    Gaze and Head & 0.730 & 0.821 & 0.775 & 0.746 \\
    Nod & 0.908 & 0.607 & 0.886 & 0.905 \\
    \bottomrule
    \end{tabular}
    \caption{Performance metrics of the \ac{eps}. Correct accuracy refers to the percentage of correct selections that were classified correctly, while incorrect accuracy indicates the percentage of incorrect selections that were classified as incorrect. Macro accuracy is the mean of correct and incorrect accuracy, and micro accuracy represents the weighted mean based on the number of samples.}
    \label{tab:EPSPerformance}
\end{table*}

Table~\ref{tab:EPSPerformance} summarizes the performance of the \ac{eps} across the three interaction methods. Overall, the EPS achieved high accuracy in classifying both correct and incorrect selections. Dwell Time and Head and Gaze showed a more balanced performance between correct and incorrect selections, whereas the Nod method had very high correct accuracy but lower accuracy for incorrect selections. Macro and micro accuracy metrics indicate that the system performs consistently across methods, with some variation in the balance between detecting correct versus incorrect selections.

\begin{figure*}
    \centering
    \includegraphics[width=\linewidth]{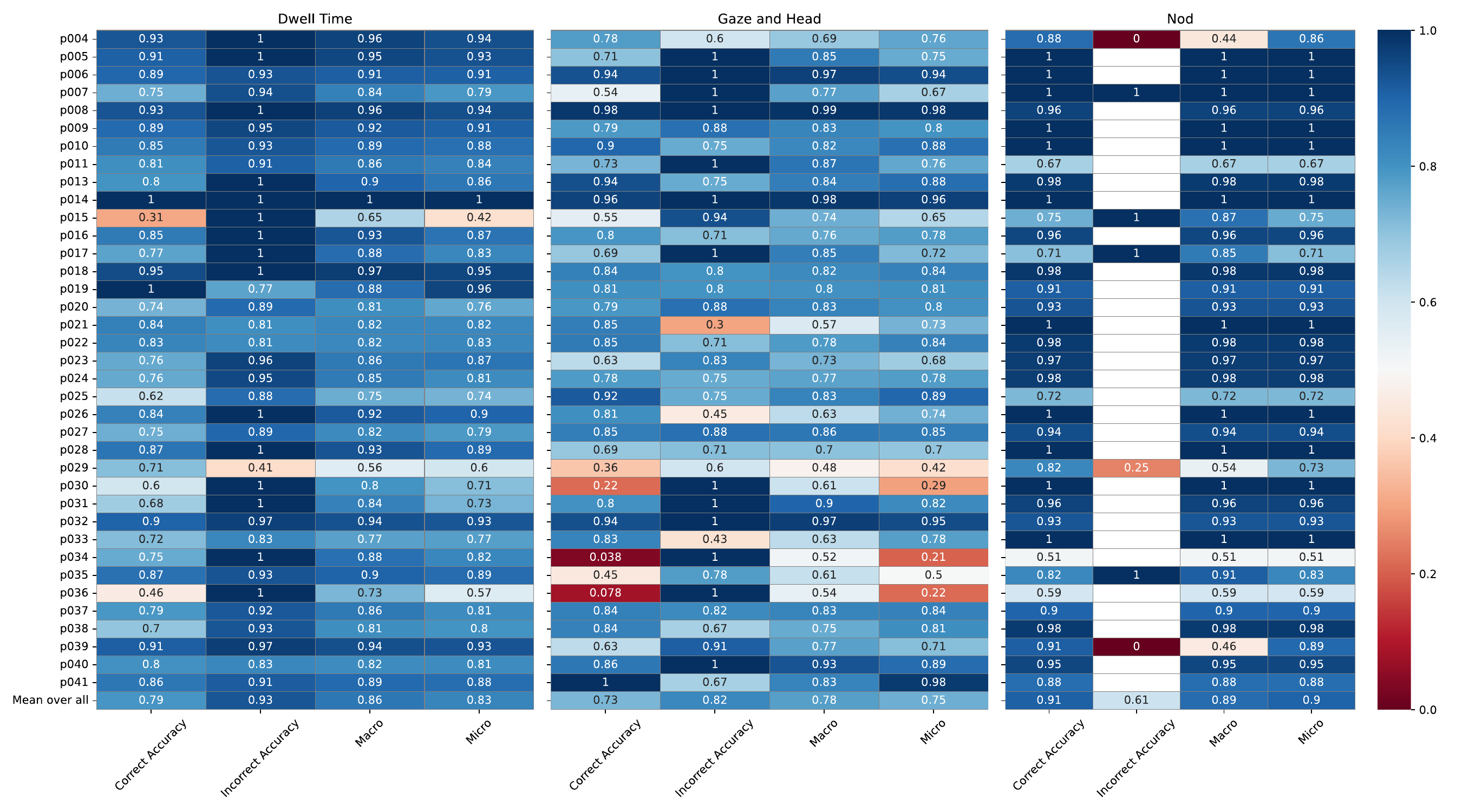}
    \caption{Individual accuracy metrics of the \ac{eps} for each participant. Blank cells indicate that no incorrect selections were made while the \ac{eps} was active. Shown metrics include correct accuracy, incorrect accuracy, macro accuracy, and micro accuracy.}
    \label{fig:EPSPerformancePerPt}
\end{figure*}

Fig.~\ref{fig:EPSPerformancePerPt} shows the performance of the \ac{eps} for each participant. Overall, the system performed well for most participants, with several achieving perfect accuracy. However, some participants exhibited lower performance, particularly in correct accuracy. The Gaze and Head method appears to have more outliers compared to Dwell Time. It is also notable that for the Nod method, only a few participants made any incorrect selections while the \ac{eps} was active.

\section{Discussion}

In this study, we developed a \ac{vr} game based on a visual search task to evaluate different interaction methods and the effectiveness of an \acf{eps}. Each participant completed multiple rounds, playing ten rounds per method (dwell time, gaze and head, and nodding), which were evenly divided between conditions with and without the \ac{eps}. After completing each method, participants filled out a questionnaire to evaluate their subjective experience. Performance was measured based on the points scored and the number of incorrect selections, while the performance of the error prevention system itself was evaluated using metrics for correct and incorrect accuracy. Below, we discuss the key findings regarding participant performance, system effectiveness, and user perception. 

The subjective results of the questionnaires show that the preferences of the participants largely corresponded to the results of our previous study without an \ac{eps}~\cite{severitt2025interplayuserpreferenceprecision}. Overall, participants rated the system positively, especially in the Dwell time and Gaze and Head conditions. In contrast, its usefulness for the Nod method was rated lower, which is due to the already low number of incorrect selections in this condition. In line with this, participants also found it more difficult to recognize whether \ac{eps} was active during Nod, whereas recognition was clearer with the other methods. Interestingly, even participants with comparatively low accuracy scores stated that they appreciated the system. One possible explanation for this is that false rejections were not perceived as disruptive, as the target could be reselected immediately while the participant's attention remained focused on it. However, it is important to maintain trust in such systems, as \citet{rittenberg2024trust} highlights that it is difficult to rebuild lost trust.

The subjective results show that \ac{eps} was generally well received and improved the participants' experience with gaze-based interaction. Participants found it particularly helpful in situations where errors occurred more frequently, while the system was perceived as less beneficial with the nod method, where errors were rare anyway. These results provide an answer to our first research question: users appreciate and trust \ac{eps} when it brings noticeable improvements without hindering interaction, but its perceived usefulness depends on the interaction method used.

\citet{peacock2022gazeinputrecognitionerror} showed that input errors can be detected within 50 to 550 ms after selection, underscoring the feasibility of rapid classification after selection. Similarly, \citet{shevtsova2025gazebasedinteractionml} showed that machine learning methods are effective for inferring interaction intentions from gaze behavior. Building on these findings, our \ac{eps} evaluates gaze behavior 200 ms after selection, and the results are consistent with participants' subjective impressions. With \ac{eps} enabled, participants made significantly fewer incorrect selections across all three interaction methods. For Dwell Time and Gaze and Head this reduction also led to higher task ratings, highlighting the system's ability to improve performance. In contrast, the benefit was minimal for the nod method, where errors were rare to begin with. Taken together, these results suggest that the \ac{eps} is particularly effective for interaction methods that are prone to incorrect selections, while its effect is limited for methods that are inherently more robust, such as Nod.

Our \ac{eps} is based on a \ac{tcn}-based architecture that has already proven itself in anomaly detection in time series data \cite{he2019tcnanomaly}. In our evaluation, the system achieved high overall accuracy, but slightly below what we had reported in previous proof-of-concept work \cite{severitt2025anomalydetection}. Correct accuracy was consistently high across all interaction methods, and incorrect accuracy was also strong—with the exception of the nod condition, where the small number of incorrect selections limited meaningful evaluation. It is noteworthy that system performance varied among participants: for some, it was nearly perfect, while for others it was less reliable. These differences point to the influence of individual viewing behavior and suggest that personal factors have a strong influence on how effectively the system can prevent errors.

The results provide an answer to our second research question: \ac{eps} is effective in a real-time environment, reducing errors and improving task performance in interaction methods where errors are likely to occur. Although its effect is smaller in methods with inherently low error rates, the system still works reliably within the time constraints of online interaction. These results confirm that anomaly detection based on gaze patterns can serve as a practical and efficient mechanism for real-time error prevention.

Overall, the results show that \ac{eps} can effectively reduce errors and improve performance in interaction methods that are more prone to incorrect selections, while also being positively received by participants. At the same time, its effectiveness is not the same for all methods or individuals. 
Furthermore, the observed differences between participants suggest that a uniform approach may not be sufficient. Previous research has shown that gaze behavior is highly individual and context-dependent \cite{hessels2020gazedifferent,gillet2021gazebehaviorrobot}, highlighting the need for more adaptive approaches. Future versions of the system could therefore benefit from dynamic adjustment of thresholds or detection strategies to better accommodate the characteristics and interaction styles of different users. Such adaptability could help ensure consistently high performance and maintain the positive user experience observed in this study.








\section{Limitations}

Although this study provides clear evidence of the effectiveness of the \ac{eps} in a controlled \ac{vr} gaming environment, there are some limitations to consider. First, the results relate specifically to the visual search task performed and may not be directly transferable to other applications or interaction contexts. In particular, the dwell time method, which worked reliably in the game, is likely to be less effective in real-world scenarios where unintended fixations and distractions from the environment are more common. Second, all evaluations were conducted within VR using the same game environment. Transferring the approach to other environments, such as alternative task types or augmented reality, may require further adjustments. In particular, a new training dataset may be needed to ensure that the system remains robust across different modalities and contexts. Future work should therefore validate the system in a broader range of environments and interaction scenarios to assess its generalizability.

\section{Conclusion}

This study demonstrates that real-time \ac{eps} based on anomaly detection can effectively support gaze-based interaction in \ac{vr} environments. Regarding the first research question, participants generally perceived the system positively and reported that it was helpful and boosted their confidence. Regarding the second research question, the system succeeded in reducing errors and improving task performance in methods prone to incorrect selections, such as Dwell Time and Gaze and Head, while the benefits were limited in methods such as nod with few errors. The variability among participants underscores the need for adaptive approaches, and future work should evaluate the system in other tasks, interaction modalities, and environments outside of \ac{vr}.


\begin{acks}
    This research is supported by European Union's Horizon 2020 research and innovation program under grant agreement No.951910 and the German Research Foundation (DFG): SFB 1233, Robust Vision: Inference Principles and Neural Mechanisms, TP TRA, project No. 276693517. We acknowledge support from the Open Access Publication Fund of the University of Tübingen.
\end{acks}

\begin{acronym}
    \acro{xr}[XR]{Extended Reality}
    \acro{vr}[VR]{Virtual Reality}
    \acro{ar}[AR]{Augmented Reality}
    \acro{mr}[MR]{Mixed Reality}
    \acro{aoi}[AOI]{Areas of Interest}
    \acro{ai}[AI]{Artificial Intelligence}
    \acro{ve}[VE]{Virtual Environment}
    \acro{nlp}[NLP]{Natural Language Processing}
    \acro{svm}[SVM]{Support Vector Machine}
    \acro{hri}[HRI]{Human-Robot Interaction}
    \acro{hci}[HCI]{Human-Computer Interaction}
    \acro{ui}[UI]{User Interface}
    \acrodefplural{UI}[UI's]{User Interfaces}
    \acro{et}[ET]{Eye Tracking}
    \acro{ae}[AE]{Auto Encoder}
    \acro{tcn}[TCN]{Temporal Convolutional Network}
    \acro{tcnae}[TCNAE]{Temporal Convolutional Network Autoencoder}
    \acro{mse}[MSE]{Mean Squared Error}
    \acro{relu}[ReLU]{Rectified Linear Unit}
    \acro{lof}[LOF]{Local Outlier Factor}
    \acro{eps}[EPS]{Error Prevention System}
\end{acronym}

\bibliographystyle{ACM-Reference-Format}
\bibliography{Content/refs}

\end{document}